\documentclass{ws-ijmpa}
\usepackage{epsfig}

\begin{document}

\markboth{Xing-Gang Wu and Tao Huang}{pion form factor}

\catchline{}{}{}{}{}

\title{Pion Electromagnetic Form Factor in the $K_T$ Factorization Formulae}

\author{Xing-Gang Wu and Tao Huang}
\address{Institute of High Energy Physics, Chinese Academy of Sciences,
P.O.Box 918(4), Beijing 100049, China}

%%%%%%%%%%%%%%%%%%%%%%%%%%%%%%%%%%%%%%%%%%%%%%%%%%%%%%%%%%%%
% You may repeat \author \address as often as necessary    %
%%%%%%%%%%%%%%%%%%%%%%%%%%%%%%%%%%%%%%%%%%%%%%%%%%%%%%%%%%%%

\maketitle

\pub{Received (received date)}{Revised (revised date)}

\begin{abstract}
Based on the light-cone (LC) framework and the $k_T$ factorization
formalism, the transverse momentum effects and the different
helicity components' contributions to the pion form factor
$F_{\pi}(Q^2)$ are recalculated. In particular, the contribution
to the pion form factor from the higher helicity components
($\lambda_1+\lambda_2=\pm 1$), which come from the spin-space
Wigner rotation, are analyzed in the soft and hard energy regions
respectively. Our results show that the right power behavior of
the hard contribution from the higher helicity components can only
be obtained by fully keeping the $k_T$ dependence in the hard
amplitude, and that the $k_T$ dependence in LC wavefunction
affects the hard and soft contributions substantially. A model for
the twist-3 wavefunction $\psi_p(x,\mathbf{k_\perp})$ of the pion
has been constructed based on the moment calculation by applying
the QCD sum rules, whose distribution amplitude has a better
end-point behavior than that of the asymptotic one. With this
model wavefunction, the twist-3 contributions including both the
usual helicity components ($\lambda_1+\lambda_2=0$) and the higher
helicity components ($\lambda_1+\lambda_2=\pm 1$) to the pion form
factor have been studied within the modified pQCD approach. Our
results show that the twist-3 contribution drops fast and it
becomes less than the twist-2 contribution at $Q^2\sim 10GeV^2$.
The higher helicity components in the twist-3 wavefunction will
give an extra suppression to the pion form factor. When all the
power contributions, which include higher order in $\alpha_s$,
higher helicities, higher twists in DA and etc., have been taken
into account, it is expected that the hard contributions will fit
the present experimental data well at the energy region where pQCD
is applicable.
\end{abstract}


%%%%%%%%%%%%%%%%%%%%%%%%%%%%%%%%%%%%%%%%%%%%%%%%%%%%%%%%%%%%
% The main text of your paper   begins here              %
%%%%%%%%%%%%%%%%%%%%%%%%%%%%%%%%%%%%%%%%%%%%%%%%%%%%%%%%%%%%

\section{Higher helicity components' contributions to the pion
form factor}

In Ref.\cite{hww}, we have systematically studied the transverse
momentum effects and the higher helicity components' contributions
to the pion form factor based on the LC framework and the $k_T$
factorization formalism\cite{lis}. The $k_T$ factorization theorem
has been widely applied to various processes and it provides a
scheme to take the dependence of the parton transverse momentum
$k_T$ into account. We note that the $k_T$ dependence in the
wavefunction can generate much larger effects than the usual
Sudakov suppression to the hard scattering amplitude in the
present experimental $Q^2$ region, which shows that it is
substantial to take the $k_T$ dependence in the wavefunction into
account.

The light cone (LC) formalism provides a convenient framework for
the relativistic description of the hadron in terms of quark and
gluon degrees of freedom, and the application of PQCD to exclusive
processes has mainly been developed in this formalism\cite{lb}. In
Ref.\cite{hww}, we have given a consistent treatment of the pion
form factor within the LC PQCD framework, i.e. both the
wavefunction and the hard interaction kernel are treated within
the framework of LC PQCD. Taking into account the spin space
Wigner rotation, one may find that there are higher-helicity
components ($\lambda_1+\lambda_2=\pm 1$) in the LC spin-space
wavefunction besides the usual-helicity components
($\lambda_1+\lambda_2=0$). The asymptotic behavior of the
hard-scattering amplitude for the higher-helicity components
including the transverse momentum in the quark propagator is of
order $1/Q^4$, which is the next to leading order contribution
compared with the contribution coming from the ordinary helicity
component, but it can give sizable contribution to the pion form
factor at the intermediate energies.

\begin{figure}
\centering \centerline{\psfig{file=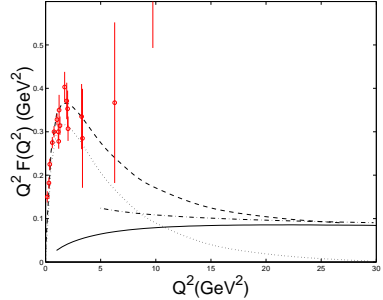,width=5cm}}
\vspace*{8pt} \caption{The combined results for the pion form
factors $Q^2F_{\pi}(Q^2)$. The solid line stands for the
contribution from the hard part, the dotted line stands for the
contribution from the soft part, the dashed line is the total Pion
form factors and the dash-dot line is the usual asymptotic
result.} \label{total}
\end{figure}

In order to compare the predictions with the experimental data,
one needs to know the contribution from the soft part. As an
example, we have considered the soft contribution to the pion form
factor with a reasonable wavefunction in the LC framework. Our
results show that the soft contribution from the higher helicity
components has a quite different behavior from that of the hard
scattering part and has the same order contribution as that of the
usual helicity ($\lambda_1+\lambda_2=0$) components in the energy
region ($Q^2\lesssim 1GeV^2$). As $Q^2>1GeV^2$, the higher
helicity components' contributions will decrease with the
increasing $Q^2$. At about $Q^2\sim 4GeV^2$, the higher helicity
components' contributions become negative and as a result the net
soft contribution to the pion form factor will then decrease with
the increasing $Q^2$, which tends to zero at about $Q^2\sim
16GeV^2$. Although the soft contribution is purely
non-perturbative and model-dependent, the calculated prediction
for the pion form factor should take the $k_T$ dependence in the
soft and hard parts into account beside including the higher order
contributions. Therefore one needs to keep the transverse momentum
in the next leading order corrections and to construct a realistic
$k_T$ dependence in the hadronic wavefunction in order to derive
more exact prediction to the pion form factor.

The combined results for the pion form factors $Q^2F_{\pi}(Q^2)$
are shown in Fig.\ref{total}, where for comparison, the
experimental data \cite{cjb} and the well-known asymptotic
behavior for the leading twist pion form factor have also been
shown.

\section{Twist-3 contributions to the pion form factor}

In Ref.\cite{hw}, we have constructed a new model wavefunction for
$\psi_p(x,\mathbf{k_\perp})$ based on the moment calculation by
using the QCD sum rule approach, i.e.
\begin{equation}\label{hel0wave}
\psi_p(x,\mathbf{k_\perp}) =(1+B_p C^{1/2}_2(1-2x)+C_p
C^{1/2}_4(1-2x))\frac{A_p}{x(1-x)}\exp \left(
-\frac{m^2+k_\perp^2}{8\beta^2x(1-x)}\right),
\end{equation}
where $C^{1/2}_2(1-2x)$ and $C^{1/2}_4(1-2x)$ are Gegenbauer
polynomials and the coefficients $A_p$, $B_p$ and $C_p$ can be
determined by the DA moments. It has a better end-point behavior
than that of the asymptotic one and its moments are consistent
with the QCD sum rule results. Although its moments are slightly
different from that of the asymptotic DA, its better end-point
behavior will cure the end-point singularity of the hard
scattering amplitude and its contribution will not be
overestimated at all.

With this model wavefunction, by keeping the $k_T$ dependence in
the wavefunction and taking the Sudakov effects and the threshold
effects into account, we have carefully studied the twist-3
contributions to the pion form factor within the modified pQCD
approach, where the model dependence of
$\psi_p(x,\mathbf{k_\perp})$ has also been given. It has been
shown that the present model of $\psi_p(x,\mathbf{k_\perp})$ can
give the right power behavior for the twist-3 contribution. This
behavior is quite different from the previous observations, where
the authors concluded that the twist-3 contribution to the pion
form factor is comparable or even larger than that of the leading
twist in a wide intermediate energy region up to $40GeV^2$. The
higher helicity components $(\lambda_1+\lambda_2=\pm 1)$ in the
twist-3 wavefunction have also been considered and it will give a
further suppression to the contribution from the usual helicity
components $(\lambda_1+\lambda_2=0)$, and at $Q^2\sim 5GeV^2$, it
will give $\sim 10\%$ suppression.

\section{CONCLUSIONS}

We have shown that both the higher helicity structure' and
twist-3's contributions are $Q^2$-suppressed to the pion form
factor. The $k_T$ factorization approach, where the transverse
momentum dependence in the wavefunction and the hard scattering
kernel have been kept, ensures these two are really power
suppressed. And both two can give sizeable contributions in the
intermediate energy regions.

\begin{figure}
\centerline{\psfig{file=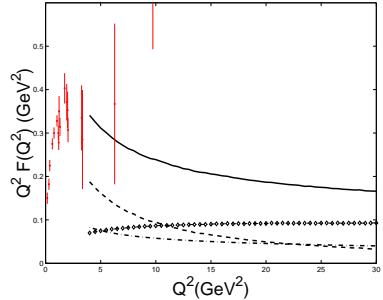,width=5cm}} \vspace*{8pt}
\caption{Perturbative prediction for the pion form factor. The
diamond line, the dash-dot line, the dashed line and the solid
line are for LO twist-2 contribution, the approximate NLO twist-2
contribution, the twist-3 contribution and the combined total hard
contribution, respectively.}\label{combine}
\end{figure}

In Fig.\ref{combine}, we show the combined hard contributions for
pion form factor, where the higher helicity components have been
included in both the twist-2 and the twit-3 wavefunctions, and the
twist-3 contribution has been calculated with the present model
wavefunction for $\psi_p(x,\mathbf{k_\perp})$ with
$\langle\xi^2\rangle=0.350$. Together with the NLO corrections to
the twist-2 contributions, which for the asymptotic DA, with the
renormalization scale $\mu_R$ and the factorization scale $\mu_f$
taken to be $\mu^2_R=\mu^2_f=Q^2$, can roughly be taken
as\cite{field},
$Q^2F^{NLO}_{\pi}\approx(0.903GeV^2)\alpha^2_s(Q^2)$, one may find
that the combined total hard contribution do not exceed and will
reach the present experimental data.

There is still a room for other power corrections, such as the
higher Fock states' contributions, soft contributions etc..
Finally, we will conclude that there is no any problem with
applying the pQCD theory including all power corrections to the
exclusive processes at $Q^2>$ a few $GeV^2$. A 12 GeV upgrade to
CEBAF will offer the possibility to measure pion form factor to
good precision out to $Q^2=36GeV^2$.  This offers the possibility
to study the transition between the dominance of `soft' and `hard'
processes in the dynamics, and to learn where the pQCD limit may
be reached, and also to check the pQCD results.

%%%%%%%%%%%%%%%%%%%%%%%%%%%%%%%%%%%%%%%%%%%%%%%%%%%%%%%%%%%%
% Doing Acknowledgement                          %
%%%%%%%%%%%%%%%%%%%%%%%%%%%%%%%%%%%%%%%%%%%%%%%%%%%%%%%%%%%%

\section*{Acknowledgements}

This work was supported in part by the Natural Science Foundation
of China (NSFC).

%%%%%%%%%%%%%%%%%%%%%%%%%%%%%%%%%%%%%%%%%%%%%%%%%%%%%%%%%%%%
% Doing references:                                %
%%%%%%%%%%%%%%%%%%%%%%%%%%%%%%%%%%%%%%%%%%%%%%%%%%%%%%%%%%%%

\end{document}